\documentclass[a4paper]{iopart}

\usepackage{iopams}     
\usepackage{graphicx}
\usepackage{epsfig}
\usepackage{color}
\usepackage{url}
\usepackage{times}
\usepackage{bm}         
\usepackage{times}

\begin{document}

\title{Constraint-preserving boundary treatment for a harmonic
  formulation of the Einstein equations}

\author{Jennifer Seiler$^{1}$, B\'{e}la Szil\'{a}gyi$^{1,2}$, Denis
Pollney$^{1}$, and Luciano Rezzolla$^{1,3}$}

\address{$^{1}$ Max-Planck-Institut f\"ur Gravitationsphysik,
  Albert-Einstein-Institut, Golm, Germany}

\address{$^{2}$ Theoretical Astrophysics,
California Institute of Technology, Pasadena, USA}

\address{$^{3}$ Department of Physics, Louisiana State University,
Baton Rouge, USA}





\date{\today}

\begin{abstract}
  We present a set of well-posed constraint-preserving boundary
  conditions for a first-order in time, second-order in space, harmonic
  formulation of the Einstein equations. The boundary conditions are
  tested using robust stability, linear and nonlinear waves, and are
  found to be both less reflective and constraint preserving than
  standard Sommerfeld-type boundary conditions.
\end{abstract}



\section{Introduction}
\label{sec:introduction}

The numerical solution of the Einstein equations has, especially in
the last two years, become a successful means of tackling many
significant physical questions. The most topical of these questions
concern the simulation of potential sources for gravitational-wave
detection, and the propagation of gravitational waves. These
problems are most commonly solved using finite-size computational
domains, and this involves imposing a boundary condition on the
physical system being simulated.

The standard treatment is to place a time-like boundary at a fixed
coordinate location, and impose boundary conditions on the dynamical
variables there. The particular conditions that are enforced ideally
satisfy a number of properties. Most importantly, in order to ensure
stability of the system, they should be compatible with the interior
evolution equations so that the discretised system forms a
\emph{well-posed} initial-boundary-value problem (IBVP). Secondly,
they should take into account the fact that Einstein evolutions 
always involve constraint equations as well as time evolution
equations, and satisfy the constraints at all times.  Otherwise,
constraint violations introduced by the boundaries are likely to drive
the evolution away from an Einstein solution.  Finally, the boundary
conditions should be compatible with physical considerations affecting
the accuracy of the solution: they should be transparent to outgoing
radiation, and restrict the amount of spurious incoming radiation from
beyond the computational domain, which is assumed to contain all of
the dynamics of interest.

To date, only the system of Friedrich-Nagy~\cite{Friedrich99}
satisfies the above conditions for the fully nonlinear vacuum
Einstein equations. In this formulation the evolution equations are
expressed first-order symmetric hyperbolic form and maximally dissipative
boundary conditions guarantee well-posedness. There have
been several attempts to approach the initial boundary value problem for
the linearized Einstein equations.  More recently, Kreiss and Winicour
proposed well-posed, constraint-preserving boundary conditions for the
linearized Einstein equations using the principle of frozen
coefficients and pseudo-differential theory of systems for the first
order system, which they then extrapolate to
second-order~\cite{Kreiss:2006mi}. Other approaches have been
suggested by Rinne for non-reflecting boundary conditions which
control incoming radiation by specifying data for the incoming fields
of the Weyl tensor~\cite{rinne-2006-23}, and by Buchman and Sarbach,
who have followed a similar route but specifying the incoming fields
at the boundary~\cite{Sarbach06}.

The approach which we introduce in this paper is partially derived
from a method first discussed in a series of related papers by Kreiss,
Winicour and collaborators~\cite{Kreiss:2006mi, Motamed06,
Babiuc:2007rr}, combined with the SBP energy method discussed in
Refs.~\cite{Mattsson2003a, Mattsson-Nordstrom-2005:SBP-operators,
Nordstrom2007a}.  By deriving energy estimates for the semi-discrete
system using the ``summation by parts'' rule (defined below), one can ensure
well-posedness~\cite{Kreiss1974a, Strand1994a,
Mattsson-Nordstrom-2005:SBP-operators, Carpenter1999a}.  By applying
this approach to boundary conditions which are radiation controlling
and constraint-preserving, we are able to construct an IBVP which
satisfies all of the above conditions in the linearized regime.  

The conditions are derived for a harmonic formulation of the Einstein
equations which has been implemented in~\cite{Szilagyi:2006qy,
Babiuc:2006wk}. The evolution equations of the formulation, given
explicitly in the next section, are first-order in time, second-order
in space. We approximate these equations using standard
finite-difference techniques, however to ensure a well-posed discrete
IBVP, we have worked out finite-difference operators for this system
which satisfy the summation by parts property. Since our
computational domain uses Cartesian coordinates on a cube, we have had
to develop consistent operators for the corners and edges, as well.
Following the developments of~\cite{Kreiss:2006mi, Calabrese02c}
and~\cite{Calabrese:2003a, kreiss-2007}, we are able to construct
boundary conditions of a Sommerfeld type, which are both well-posed
and satisfy both the Einstein and Harmonic constraints.

We have used the newly constructed boundary conditions in a number of
practical tests and found them to perform extremely well in comparison
with other standard techniques.  Test evolutions include linear and
nonlinear waves. In each case, the new boundary conditions are found
to be more transparent to outgoing waves, as well as reducing the
overall constraint violations on the grid.  Further, the evolutions
are stable against perturbations by high-frequency constraint
violation (``noise'') added to the data, providing a strong
demonstration of their robustness.  Tests were also done for black
hole space-times. For head-on collisions and inspiral, our boundary
conditions showed improvements in reducing reflections and constraint
preservation, and thus improved the waveform accuracy, but as the
standard treatment was not long term stable for these tests due to
instabilities at the excision boundary, we did not feel that it was
appropriate to display in our results.

The plan of the paper is as follows: in the following section we
introduce our harmonic evolution system, and describe the evolution
variables, equations, and constraints.  Section~\ref{sec:harmonic}
describes the implementation of a harmonic evolution system for the
Einstein equations. In particular, Section~\ref{sec:fd} describes the
construction of finite-difference operators which ensure that the
discretisation of our evolution equations remains well-posed,
including the boundary faces, corners and edges.  The boundary
treatment is described in Section~\ref{sec:boundary_treatment}. In
Section~\ref{sec:cpbc} we present the derivation of constraint
preserving Sommerfeld type conditions for this system. Finally, in
Section~\ref{sec:results}, we discuss a number of test cases to which
these boundary conditions have been applied, and demonstrate their
usefulness in ensuring stability and improving accuracy in a variety
of scenarios.

\section{The harmonic evolution system}
\label{sec:harmonic}

\subsection{Formulation of the evolution equations}

The decomposition of the Einstein tensor into evolution equations and
constraints leaves four degrees of freedom in the space-time metric
that are not set by the field equations themselves, but can be freely
specified.  In a $3+1$ approach, these four degrees of freedom are
determined by the choice of the lapse and shift, which amounts to
specifying four out of ten metric components. The
Arnowitt-Deser-Misner (``ADM'') equations~\cite{Arnowitt62} are a well
known reduction of the Einstein system corresponding to this style of
gauge choice.

An alternate approach to fixing the gauge degrees of freedom specifies
the action of the wave operator on the coordinates, regarded as four
scalar quantities.  This is done by first choosing four functions
$F^\alpha$ and then constructing a coordinate map $x^\alpha$ subject
to the condition~\cite{Friedrich85} that the d'Alembertian of each
coordinate is 
\begin{equation}
  \Box x^{\alpha} = \frac{1}{\sqrt{-g}}\partial_{\mu}\left(\sqrt{-g}g^{\mu\beta}\partial_{\beta} x^{\alpha}\right) = F^{\alpha}\,,
\label{eq:gaugechoice}
\end{equation}
rewriting Eqs.~(\ref{eq:gaugechoice}) as constrained variables
\begin{equation}
C^\alpha := \Box x^\alpha - F^\alpha = 0\, ,
\label{eq:gaugechoice2}
\end{equation}
and using them in combination with the Einstein tensor
$G^{\mu\nu}$
one obtains the generalized harmonic evolution system
\begin{eqnarray}
E^{\mu\nu} := G^{\mu\nu} 
- \nabla^{(\mu} C^{\nu)}
+ \frac{1}{2} g^{\mu\nu} \nabla_\alpha C^{\alpha} = 0\,.
\label{eq:reducedeinstein}
\end{eqnarray}
In terms of these variables, the vacuum Einstein equations are a
system of ten wave equations acting on the metric components, coupled
through the coefficients of the wave operator and the source terms.

Using the densitized inverse metric
$\tilde g^{\mu\nu} := \sqrt{-g} g^{\mu\nu}$ as evolution variables,
the harmonic constraints~(\ref{eq:gaugechoice2}) take the form
\begin{equation}
  C^{\alpha}= -\frac{1}{\sqrt{-g}}\partial_{\beta}\tilde{g}^{\alpha\beta}
    - F^\alpha = 0\,,
  \label{eq:constraints}
\end{equation}
while for the evolution equations we obtain
\begin{eqnarray}
\fl
\partial_\rho \left( g^{\rho\sigma} \partial_\sigma {\tilde g}^{\mu\nu} \right)
      - 2\sqrt{-g} g^{\rho\sigma} g^{\tau \lambda}
        \Gamma^\mu_{\rho\tau} \Gamma^{\nu}_{\sigma \lambda}
        - \sqrt{-g} (\partial_\rho g^{\rho\sigma}) (\partial_\sigma g^{\mu\nu})
        + \frac{g^{\rho\sigma}}{\sqrt{-g}} (\partial_\rho g^{\mu\nu})
                (\partial_\sigma g)
&& \nonumber\\
+ \frac{1}{2} g^{\mu\nu}
          \bigg (
   \frac{g^{\rho\sigma} }{2g\sqrt{-g}} (\partial_\rho g) (\partial_\sigma g)
   + \sqrt{-g} \Gamma^\tau_{\rho\sigma} \partial_\tau g^{\rho\sigma}
+ \frac{1}{\sqrt{-g}}(\partial_\sigma g)
            \partial_\rho  g^{\rho\sigma}     \bigg )
&& \nonumber \\
        + 2 \sqrt{-g} \nabla^{(\mu} F^{\nu)}
        - \sqrt{-g} g^{\mu\nu} \nabla_\rho F^\rho
        + \sqrt{-g} A^{\mu\nu}
                = 0 \,,
\label{eq:harmonic_full}
\end{eqnarray}
where in the final term we have allowed for a constraint
adjustment function which may depend on the metric
and its first derivatives,
\begin{equation}
A^{\mu\nu}:= C^\rho A^{\mu\nu}_\rho
     (x^\rho,g_{\rho\sigma},\partial_\tau  g_{\rho\sigma}) \,.
\end{equation}
The constraint adjustment implemented in the code are given
from~\cite{Babiuc-etal-2005} and have the form
\begin{equation}
A^{\mu\nu}:= -\frac{a_{1}}{\sqrt{-g}}C^{\rho}\partial_{\rho}\tilde{g}^{\mu\nu}+ 
\frac{a_{2}C^{\rho}\nabla_{\rho}t}{\varepsilon + \epsilon_{\sigma\tau}C^{\sigma}C^{\tau}}
C^{\mu}C^{\nu} - \frac{a_{3}}{\sqrt{-g^{tt}}}C^{(^{\mu}\nabla^{\nu})}t \, ,
\end{equation}
where the $a_{i} > 0$ are adjustable parameters,
$\epsilon_{\sigma\tau}$ is the natural metric associated with the
Cauchy slicing, and $\varepsilon$ is a small positive number chosen to
ensure regularity.

Assuming that the gauge source functions $F^\alpha$ are also chosen
such that they do not depend on derivatives of the metric, then the
principle part of Eq.~(\ref{eq:harmonic_full}) consists of only its
first term. That is, we have a set of ten wave equations of the form
\begin{equation}
\partial_\rho \left( g^{\rho\sigma} \partial_\sigma {\tilde g}^{\mu\nu} \right)
 = S^{\mu\nu} \,,
\label{eq:harmonic_concise}
\end{equation}
where $S^{\mu\nu}$ are non-principle source terms consisting of at
most first derivatives of the evolution variables.  By implication,
this system inherits the property of the well-posedness of the
initial-boundary value problem for the wave equation.

It is essential to have all of the initial data constructed in a way
that satisfies the conditions
\begin{equation}
C^\rho = 0\,, \qquad \partial_t C^\rho = 0\,,
\end{equation}
as well as a construction of the boundary data that implies a
homogeneous boundary condition for the constraints. However, by
satisfying these conditions, we arrive at a well-posed IBVP for the
constraint propagation system. Recent work by Kreiss, Winicour, Reula
and Sarbach~\cite{kreiss-2007} demonstrates that it is possible to
construct such boundary data while keeping the IBVP of the evolution
system of the metric variables well-posed. In the following sections
we implement and test such boundary conditions and compare them with
simpler (unconstrained SAT and non-SAT) boundary treatments for a
number of test-problems.  

In order to understand the feasibility of Eq.~(\ref{eq:harmonic_full})
as an unconstrained evolution system, one needs to have insight into
the associated constraint propagation
system~\cite{Gundlach2005:constraint-damping, Brodbeck98, Reula98a}
\begin{equation}
\square C^{\rho} = S^\rho (g, \partial g, \partial^2 g, C, \partial C,
A, \partial A)\,,
\label{eq:constrainteq}
\end{equation}
where $S^\rho$ is a source term dependent on the metric, the
constraints, the constraint adjustment term, and their derivatives.

The principal part of Eq.~(\ref{eq:constrainteq}) is, again, that of a
wave operator, implying the connection to results regarding the
well-posedness of the initial-boundary value problem of the wave
equation.

We have implemented the generalized harmonic evolution
system~(\ref{eq:harmonic_full}), cast in a form that is first
differential-order in time, and second-differential order in space.
The auxiliary variables
\begin{equation}
  Q^{\mu\nu} \equiv n^{\rho}\partial_{\rho}\tilde{g}^{\alpha\beta} \,,
  \label{eq:Qdef}
\end{equation}
are used to eliminate the second time-derivatives, where $n^{\rho}$ is
time-like and tangential to the outer boundary~\cite{Szilagyi:2006qy}.
The resulting evolution system takes the form
\begin{eqnarray}
\fl
&& \qquad \partial_{t}\tilde{g}^{\mu\nu} = -\frac{g^{it}}{g^{tt}}
    \partial_{i}\tilde{g}^{\mu\nu} + \frac{1}{g^{tt}}Q^{\mu\nu}\,,
  \label{eq:dtg}
 \\
\fl
&&\qquad  \partial_{t}Q^{\mu\nu} = -\partial_{i}\left(\left(g^{ij} 
    - \frac{g^{it}g^{jt}}{g^{tt}}\right)  \partial_{j}\tilde{g}^{\mu\nu}\right)
    - \partial_{i}\left(\frac{g^{it}}{g^{tt}}Q^{\mu\nu}\right) 
    + \tilde{S}^{\mu\nu}(\tilde{g},\partial\tilde{g}, F,\partial F)\,,
  \label{eq:dtQ}
\end{eqnarray} 
where $\tilde{S}^{\mu\nu}(\tilde{g},\partial\tilde{g}, F,\partial F)$
are non-principle source terms consisting of at most first derivatives
of the evolution variables and are determined by our choice of gauge.

\subsection{Discretisation and finite differencing}
\label{sec:fd}

The numerical implementation of~(\ref{eq:dtg}-\ref{eq:dtQ}) follows
the ``method of lines'' approach, which applies to systems
which can be cast in the form of an ordinary differential equation
containing some spatial differential operator $\mathbf{L}$
\begin{equation}
  \partial_t \mathbf{q} = \mathbf{L}(\mathbf{q}).
\end{equation}
The time integration can be carried out using standard methods, such
as the Runge-Kutta algorithm.

Spatial derivatives on the right-hand-sides
of~(\ref{eq:dtg}-\ref{eq:dtQ}) are computed by finite differencing on
a uniformly spaced Cartesian grid.  We have implemented finite
difference stencils which are fourth-order accurate over the interior
grid and second-order accurate at the boundaries. To ensure
well-posedness of the semi-discrete system, we need to obtain an
estimate on the energy growth of the system.  To do this, we have used
difference operators $D$ which satisfy the ``summation by parts''
(SBP) property.  A discrete operator is said to satisfy SBP for a
scalar product $E=\langle u,v\rangle = \int_{a}^{b}u\cdot vdx $ if
\begin{equation}
\label{eq:sbp}
\langle u, Dv \rangle + \langle v, Du\rangle = (u\cdot v) \mid_{a}^{b}\,,
\end{equation}
holds for all functions $u$, $v$ in the domain $[a,b]$. This is the
discrete analog of the integration by parts property of continuous
functions. By integrating for our energy estimate using the SBP
property of our difference operators, we ensure that boundedness
properties of the continuum energy estimate carry over to the
discretised system. We can construct these difference operators,
including numerical boundary conditions in a consistent way, for the
system of equations in~(\ref{eq:harmonic_concise}).

We follow the procedure outlined by Strand~\cite{Strand1994a} in
constructing finite difference stencils $D$ of a given order, $\tau$,
such that
\begin{equation}
Du=\frac{du}{dx} + \mathcal{O}(h^{\tau}),
\end{equation}
and which satisfy the SBP property (\ref{eq:sbp}). Briefly, 
given a state vector $u = (u_0,u_1,\ldots, u_n)^T$ on $n$ grid
points, we construct a finite difference operator $D$ as a matrix
acting on $u$. The coefficients of $D$ can be represented as products
of the standard operators
\begin{eqnarray}
D_{0x}f_{i,j,k}=\frac{1}{2h}\left(f_{i+1,j,k}-f_{i-1,j,k}\right)\,,
\nonumber \\ 
D_{+x}f_{i,j,k}=\frac{1}{h}\left(f_{i+1,j,k}-f_{i,j,k}\right)\,,
\nonumber \\ 
D_{-x}f_{i,j,k}=\frac{1}{h}\left(f_{i,j,k}-f_{i-1,j,k}\right)\,.
\end{eqnarray}
They are determined up to the boundaries of the domain by solving the
set of polynomials
\begin{equation}
Dx^{m}-\frac{dx^{m}}{dx} = 0, \qquad m=0,1,\ldots,\tau,
\end{equation}
which establish the order of accuracy $\tau$ of the approximation.
The SBP rule~(\ref{eq:sbp}) provides an additional set of restrictions,
\begin{equation}
 \langle u,Du\rangle  = -\frac{1}{2}u^{2}(0)\,,
\end{equation}
and
\begin{equation}
 \langle u+v,D\left(u+v\right)\rangle_{h}  = 
   \langle D\left(u+v\right), u+v \rangle_{h} - \left(u_{0}+v_{0}\right)^{2}\,,
\end{equation}
which should hold for all $u, v$ in the half line divided into intervals of
length $h > 0$. Following Strand~\cite{Strand1994a}, we can solve these
conditions explicitly for the stencil coefficients of the first
derivative operator $D$. It is trivial to obtain a second derivative
operator simply by repeated application of the derived first derivative
operator. However, this results in a very wide and thus impractical
stencil, and instead we use the second derivative SBP operators
described in~\cite{Carpenter1994a, Carpenter1999a}. The explicit
expressions for the finite difference stencils which we use are
given in~\cite{Szilagyi:2006qy}.

The above considerations apply to the construction of difference
operators along a single coordinate direction. We can derive a
3D SBP operator by applying the 1D operator along each coordinate
direction.  It can be shown that the resulting operator
also satisfies SBP with respect to a diagonal scalar product
\begin{equation}
  \langle u,v\rangle_{H} = h_{x}h_{y}h_{z}\sum_{ijk}{}
    \sigma_i\sigma_j\sigma_k u_{ijk}\cdot v_{ijk},
\end{equation}
where $\sigma_i$, $\sigma_j$, $\sigma_k$ are the coefficients of the
corresponding inner product in each of the coordinate directions. The
norm $H$ is defined such that for a discrete inner product $\langle
u,v\rangle_{H} = u^{T}Hv$, where $H=H^{T}>0$. Note that this is only true
if the norm, $H$, is diagonal. Here we restrict ourselves to this case.

\section{Boundary Treatment}
\label{sec:boundary_treatment}

\subsection{Well-posed Boundary Conditions}
\label{sec:bc} 

We have constructed finite differencing operators which satisfy
summation by parts, and thus can use the rule (\ref{eq:sbp}) as a tool
for deriving an energy estimate and ensuring well-posedness of the
semi-discrete system. For the continuum system, we have a well defined
energy estimate which can be used to bound solutions. Through use
of the SBP-compatible derivative operators defined in the previous
section, we ensure that an energy estimate also holds for the
semi-discrete system. If this energy estimate bounds the norm of the
solution in a resolution independent way, then we have a stable
semi-discrete system.  Optimally, we would like the norm of the
semi-discrete solution to satisfy the same estimate as the continuum
solution.

To establish well-posedness we impose boundary conditions based 
upon the energy norm
\begin{equation}
  \mathcal{E} = \Vert u(t,.) \Vert ^{2} = \langle u, u\rangle 
    = \int_{\Omega}{u\cdot Hudx}
\end{equation}
where $u(t,.)$ is the solution of the IBVP at time t, and $H$ is a 
symmetric positive definite matrix on the bounded domain $\Omega$.  
We require that
\begin{equation}
  \mathcal{E}(t) \leq C(t)\mathcal{E}(0)\,,\qquad t\geq 0\, ,
  \label{eq:Ebound}
\end{equation}
with $C(t)$ independent of the initial and boundary data, so that
the solution is bounded by the energy at time $t=0$ for all $t$.

As an instructive example which contains the essential features of the
derivation for the Einstein equations, we derive explicitly the
energy estimate for the wave equation with shift in the Appendix.

We require that the energy, $\mathcal{E}^{\left(n\right)} =
\Vert u\left(\cdot,t\right)\Vert^{2},$ satisfies~(\ref{eq:Ebound}) for
positive times, that is, for the duration of a simulation the energy
is bounded. The use of simultaneous approximation terms (the SAT or
'penalty') allows us to choose values for the free parameters
in the boundary terms which conserve the energy in the system. We
determine the time dependence of the energy for this system in order
to derive coefficients for our penalty terms at the boundary points
which give a well-posed semi-discrete system. For the wave equation
with shift the semi-discrete evolution equations which results from the
SBP-SAT calculation in the appendix (Sec.~\ref{sec:appendix}) are
\begin{eqnarray}
\fl
\qquad
   u_{tt} & = & -\frac{\gamma^{it}}{\gamma^{tt}}H^{-1}D^{(1)}_{i}u_{t}
  - \frac{\gamma^{ij}}{\gamma^{tt}}H^{-1}D^{(2)}_{ij}u 
  - \frac{\gamma^{ij}}{\gamma^{tt}\beta_{0_{i}}}H^{-1}E_{0_{i}}
  (\alpha_{0_{i}}u_{t} + \beta_{0_{i}}S_{i}u  + \delta_{0_{i}}u) 
  \nonumber\\
\fl
\qquad
  && +\frac{\gamma^{ij}}
  {\gamma^{tt}\beta_{N_{i}}}H^{-1}E_{N_{i}}(\alpha_{N_{i}}u_{t}
  +\beta_{N_{i}}S_{i}u + \delta_{0_{i}}u )\label{eq:finalsbp}\, ,
\end{eqnarray}
where $D^{(1)}_{i}$ is the discrete first derivative operator in the
$i$ direction and $D^{(2)}_{ij}$ is the discrete second derivative
operators which are blended to sideways differencing near the
boundaries.  This equation, as a result of the application of the SAT
terms, satisfies the energy conservation equation $d\mathcal{E}/dt =
0$. The corresponding calculation for the Einstein equations,
Eq.~(\ref{eq:dtg}--\ref{eq:dtQ}) mirrors this calculation, except with
the inclusion of source terms which do not themselves modify the
boundary treatment.

\subsection{Constraint preservation}
\label{sec:cpbc}

In ref.~\cite{Szilagyi:2006qy}, we used a somewhat ad-hoc boundary
condition, which applies a Sommerfeld-like dissipative operator to all
ten components of the metric
\begin{equation}
  \left(\partial_t + \partial_x - \frac{1}{r}\right) (g^{\mu\nu} - g^{\mu\nu}_0) = 0\, .
  \label{eq:sommerfeld}
\end{equation}
This follows the physically motivated reasoning that far away from a
source, the evolution variables each satisfy a generally radial
outgoing wavelike behaviour. The condition is particularly simple to
apply, and has been used extensively in evolutions using a
conformal-traceless formulation of the Einstein equations~(see, for
example, \cite{Pollney:2007ss}), where the choice of evolution
variables has so far hindered the development of a more rigorous
boundary treatment. In fact, in simulations where the boundaries have
been pushed to large distances (for instance through the use of mesh
refinement), the condition has proven to be useful enough to allow for
long-term stable evolutions. Eventually, however, boundary effects do
contaminate the interior grid, and can lead to a loss of convergence
or the accuracy required to resolve delicate physical features. The
conditions given by Eq.~(\ref{eq:sommerfeld}) make no effort to
satisfy the Einstein constraints, and thus can over time drive the
solution away from a solution of the full Einstein equations.

For the Einstein equations in harmonic form, it is possible to derive
consistent boundary conditions by explicitly evaluating the constraint
propagation system. This has been done for the first order harmonic
evolution system described by Lindblom et al.~\cite{Lindblom:2005qh},
who have derived consistent conditions based on limiting incoming
characteristics.

Alternatively, Kreiss and Winicour~\cite{Kreiss:2006mi} have
demonstrated a set of Sommerfeld type boundary conditions, which
are strongly well posed, as well as preserving the harmonic
constraints. The well-posedness follows from results in
pseudo-differential theory of strongly well-posed systems, and
applies to a broad class of conditions. We can apply their results
directly to the generalized harmonic evolution system used here.
The harmonic constraints, Eq.~(\ref{eq:constraints}), provide
conditions for the time components of the metric:
\begin{equation}
  -\partial_{t}g^{\mu t} - \partial_{x}g^{\mu x}
   - \partial_{y}g^{\mu y} - \partial_{z}g^{\mu z} - F^{\mu}
  = 0\,.
\end{equation}
The remaining metric components are determined by applying
the Sommerfeld-type condition, Eq.~(\ref{eq:sommerfeld}), in
a hierarchical fashion, using previously determined components
as required:
\begin{eqnarray}
&&  \left(\partial_x + \partial_t + \frac{1}{r}\right) \left(g^{AB} 
    - g^{AB}_{0}\right) = 0\, , \\
&&  \left(\partial_x + \partial_t + \frac{1}{r}\right) \left(g^{tA}
    - g^{xA} - g^{tA}_{0} + g^{xA}_{0}\right) = 0\, , \\
&&  \left(\partial_x + \partial_t + \frac{1}{r}\right) \left(g^{tt}
    - 2g^{xt} + g^{xx}
    -g^{tt}_{0} + 2g^{xt}_{0} - g^{xx}_{0}\right) = 0 \, .
\end{eqnarray}
These particular conditions are chosen to ensure well-posedness of the
solution, but are not unique. They lead to the following explicit
conditions on the positive $x$ boundary:
\begin{eqnarray}
  \left(\partial_x + \partial_t \right) g^{0\mu} & = &
  \partial_{x}g^{0\mu} - \partial_{x}g^{1\mu}
  - \partial_{y}g^{2\mu} - \partial_{z}g^{3\mu} - F^{\mu}\,, \label{eq:cpb1}\\
  \left(\partial_x + \partial_t \right) g^{11} & = &
  \left(\partial_x + \partial_t \right) \left(2g^{01} -g^{00}\right)
  - \frac{1}{r}\left(g^{11} -2g^{01}+g^{00}\right)
  \label{eq:cpb2} \nonumber\\
  && + \left(\partial_x + \frac{1}{r}\right) \left(g^{11}_{0}-2g^{01}_{0} +g^{00}_{0}\right) \,,\\
 \left(\partial_x + \partial_t \right) g^{1A} & = &
  \left(\partial_x + \partial_t \right) \left(g^{0A} 
  - g^{0A}_{0}\right) \label{eq:cpb3}\nonumber\\
&&  - \frac{1}{r} \left(g^{1A} - g^{1A}_{0}\right) 
  + \frac{1}{r} \left(g^{0A} 
  -g^{0A}_{0}\right) - \partial_x g^{1A}_{0}\,,\\
  \left(\partial_x + \partial_t \right) g^{AB} 
  & = & - \frac{1}{r} \left(g^{AB} -g^{AB}_{0}\right)
  + \partial_x g^{AB}_{0}\,.
  \label{eq:cpb4}
\end{eqnarray}

We combine the results of the previous section (see Appendix) with
these constraint preserving conditions, to arrive at expressions for
the evolution equations for $Q^{\mu\nu}$ from Eq.~(\ref{eq:dtQ}) with
the new penalties derived in the appendix and shown in
Eq.~(\ref{eq:finalsbp}),
\begin{eqnarray}
  \partial_{t}Q^{\mu\nu} & = &
  - \left(g^{ij}+\frac{g^{it}g^{jt}}{g^{tt}}\right)D_{i\pm}D_{j\mp}\tilde{g}^{\mu\nu} 
  -\frac{g^{it}}{g^{tt}}D_{i}Q^{\mu\nu} + \tilde{S}^{\mu\nu}
  \nonumber\\
  && +\frac{2g^{ij}}{g^{tt}\beta_{0}}H^{-1}E_{0_{i}}\left[\left(1+\frac{g^{it}}{g^{tt}}\right)\tilde{g}^{\mu\nu}_{t}
  + S_{i+}\tilde{g}^{\mu\nu} - p^{\mu\nu}\right]
  \nonumber\\
  &&+\frac{2g^{ij}}{g^{tt}\beta_{N}}H^{-1}E_{N_{i}}\left[\left(1-\frac{g^{it}}{g^{tt}}\right)\tilde{g}^{\mu\nu}_{t}
  + S_{i-}\tilde{g}^{\mu\nu} - p^{\mu\nu}\right]\,,
\end{eqnarray}
where the $p^{\mu\nu}$ are determined by Eqs.~(\ref{eq:cpb1})--(\ref{eq:cpb4}).
For example
\begin{equation}
p^{0\mu}= S_{i+}\tilde{g}^{0\mu}- (S_{i+}\tilde{g}^{i\mu} + D_{A+}\tilde{g}^{\mu A}  + D_{B+}\tilde{g}^{\mu B} + F^{\mu})\,,
\end{equation}
corresponds to the constraint conditions in Eqs.~(\ref{eq:cpb1}),
where $i$ is the direction outward from the boundary face, $S_{i\pm}$
is the stencil for sideways finite differencing on the boundary, and
$A$, and $B$ are tangent to the face.

\section{Applications}
\label{sec:results}
\begin{figure}
  \begin{center}
    \resizebox{85mm}{!}{
      \includegraphics{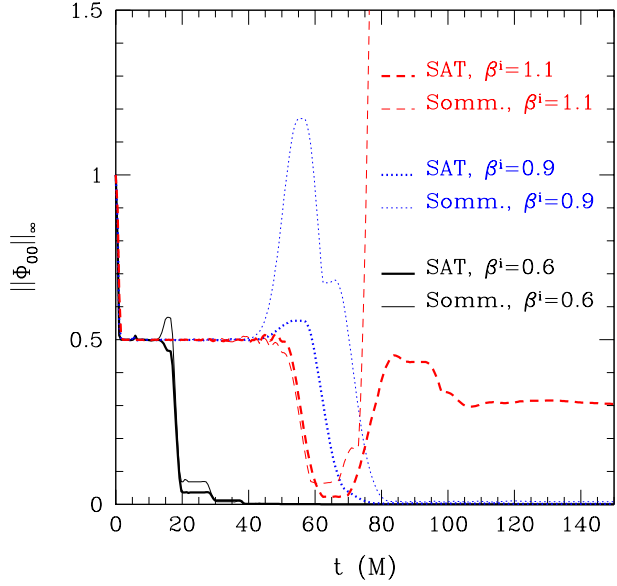}
    }
    \caption{The evolution of $\phi$ for flat-space wave equations with a
      constant shift in the $x$-direction.  As initial data we have used
      a spherical Gaussian pulse of amplitude $1.0$ and width $1.0$, on a
      grid $8$ ($121$ grid points) units in size.  Thin lines are the
      Sommerfeld-type boundary conditions without the SAT terms applied,
      whereas thick lines use the SAT boundary treatment given by
      Eq.~(\ref{eq:sat}).  \label{fig:shifted_wave_a}}
  \end{center}
\end{figure}
\begin{figure}
  \begin{center}
    \resizebox{85mm}{!}{
      \includegraphics{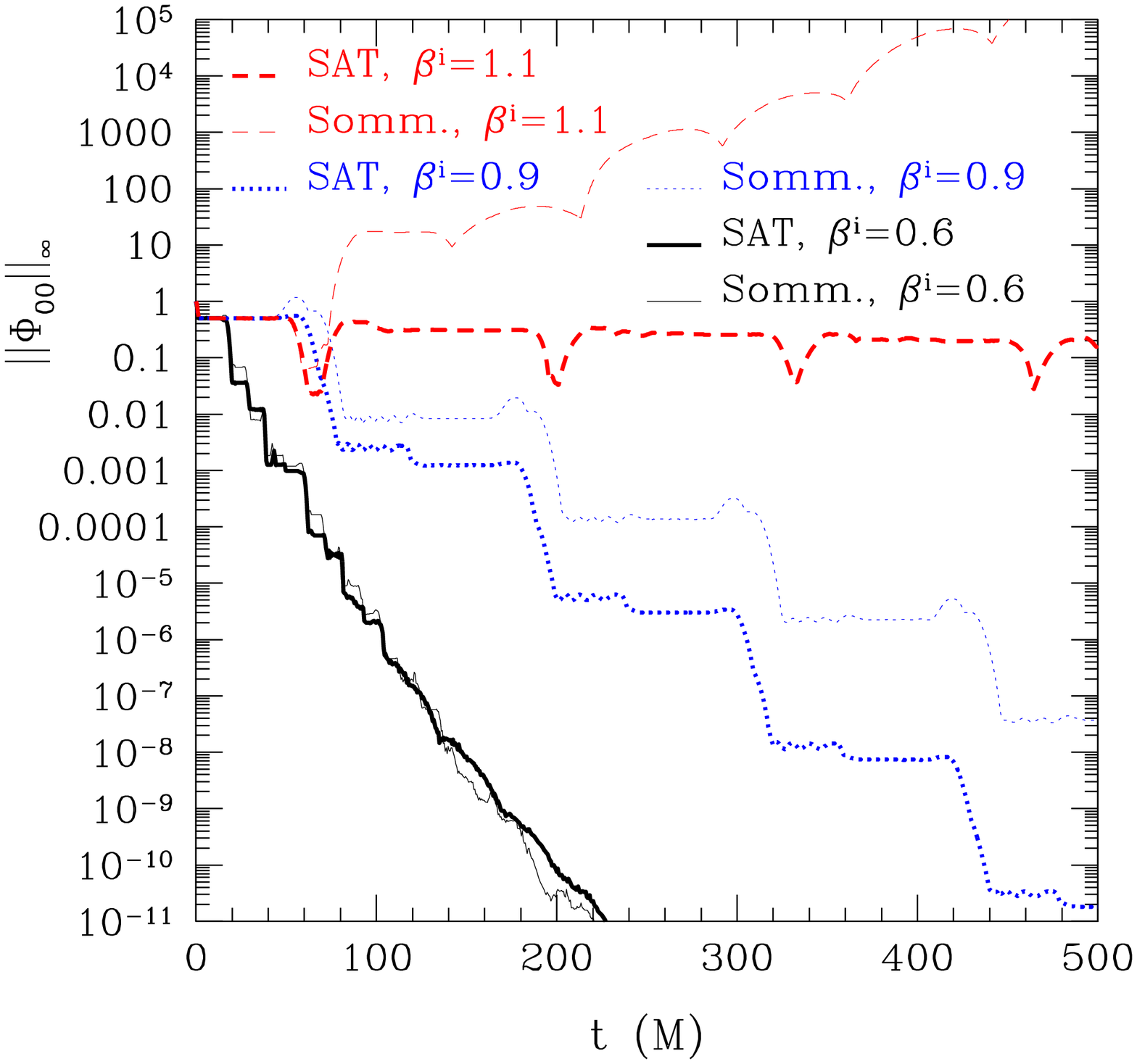}
    }
    \caption{The same as in Fig.~\ref{fig:shifted_wave_a} but shown in a
      logarithmic scale for $||\Phi_{00}||_{\infty}$ and on a longer
      timescale. Note that standard Sommerfeld boundary conditions are
      unstable for $|\beta^i| > 1$.  \label{fig:shifted_wave_b}}
  \end{center}
\end{figure}

The boundary prescription described in the previous section has been
implemented for our harmonic Einstein evolution code (\cite{Szilagyi:2006qy}
and Sec.~\ref{sec:harmonic}). We have carried out tests comparing
three boundary configurations. The first, which we refer to as
``standard Sommerfeld'' simply applies Eq.~\ref{eq:sommerfeld} to
each evolution variable on each face of the cubical evolution domain,
which was the boundary implementation used in~\cite{Szilagyi:2006qy}.
The second (``SAT'') applies the boundary treatment derived in
Sec.~\ref{sec:bc}, and the third (``CP-SAT'') improves on this by
implementing the constraint preserving conditions of Sec.~\ref{sec:cpbc}.
We find that in each case, the SAT and CP-SAT boundary conditions respectively
improve on the standard Sommerfeld condition in their ability to reduce
boundary reflections and constraint violations over time.

\subsection{Shifted waves}
\label{sec:shift}

As a first test of the methodologies outlined in the previous
section, we consider a simplified non-relativistic example
problem which demonstrates the effectiveness of the SAT
method. One of the challenges of designing boundary treatments which
control the energy growth for black hole space-times in commonly
used gauges is the problem of non-zero shift. A useful
problem which has been used as a toy model for the full
Einstein equations is the shifted scalar wave 
equation~\cite{Calabrese:2005fp, Babiuc-etal-2005},
\begin{equation}
\left( \partial_{t}^{2} - 2\beta^{i}\partial_{i}\partial_{t} 
   - \left( \delta^{ij} - \beta^{i}\beta^{j}\right)
      \partial_{i}\partial_{j}\right) \phi = 0\, ,
  \label{eq:shifted_wave}
\end{equation}
with shift vector $\beta^i = g^{it}/g^{tt}$ (see
Eq.~(\ref{eq:shwave})). In the appendix, we have explicitly derived the
boundary treatment of this problem, which has been implemented in a 3D
evolution code.

In Fig.~\ref{fig:shifted_wave_a}, we display results from
evolutions of a Gaussian wave packet, for various constant values of
the shift. The $L_\infty$-norm of the energy of the solution is
plotted as a function of time for evolutions using standard Sommerfeld
type conditions, Eq.~(\ref{eq:sommerfeld}), and compared with the SAT
conditions derived in Sec.~\ref{sec:bc}.  As the waveform impinges on
the boundary, there is a certain amount of unphysical reflection, but
the energy is largely removed from the grid in steps corresponding to
the crossing time, as visible in Fig.~\ref{fig:shifted_wave_b}.  The
boundary reflections are much lower in the case of the SAT boundary
conditions, and the evolution is stable even to superluminal,
$|\beta^i| > 1$, shifts suggesting that our conditions are stable even
for outflow boundaries.

\begin{figure}
  \begin{center}
    \resizebox{85mm}{!}{
      \includegraphics{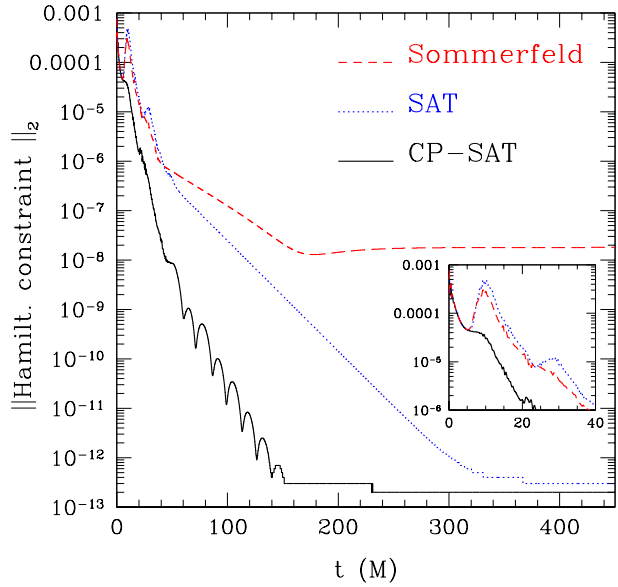}
    }
    \caption{The $L_2$-norm of the Hamiltonian constraint for a
      Teukolsky wave, comparing our constraint-preserving boundary
      conditions with the standard non-SBP Sommerfeld conditions, as
      well as the purely Sommerfeld SAT algorithm to ensure
      well-posedness. \label{fig:teukolsky}}
  \end{center}
  
\end{figure}
\subsection{Linear waves}
\label{sec:teukolsky}
\begin{figure}[h]
  \begin{center}
      \includegraphics{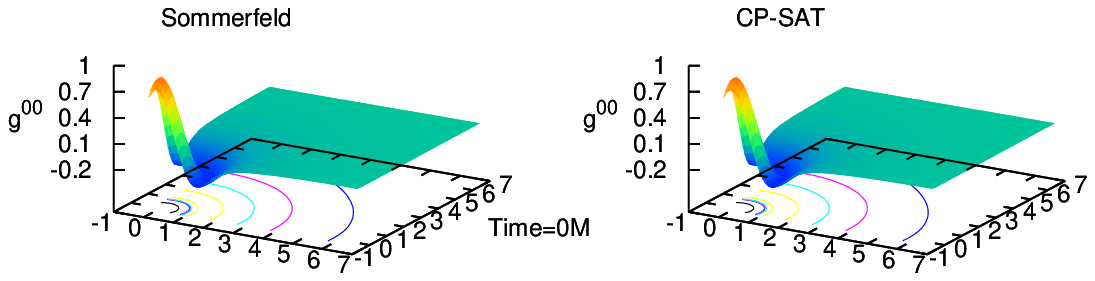}\\
      \hspace{2pc}\includegraphics{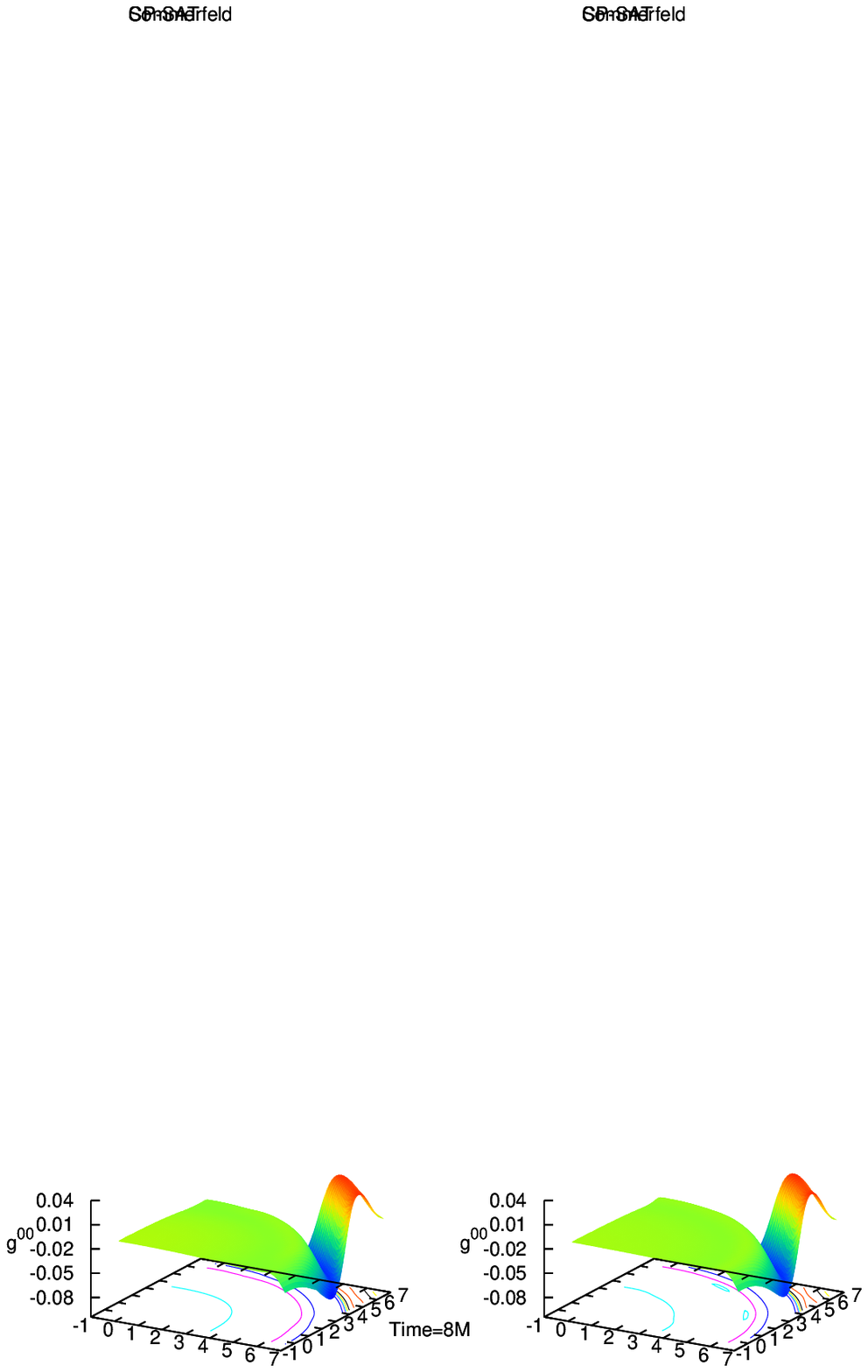}\\
      \hspace{2pc}\includegraphics{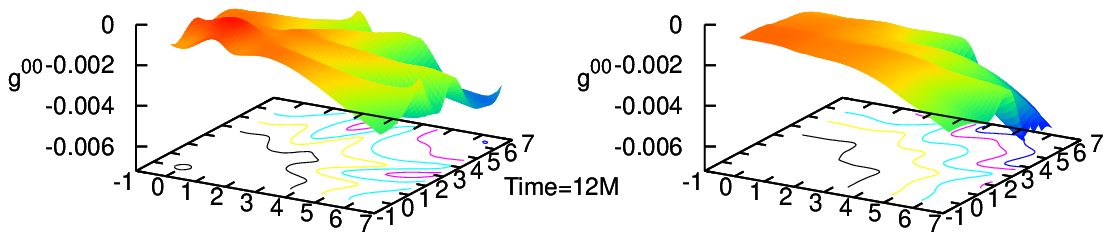}\\
      \hspace{2pc}\includegraphics{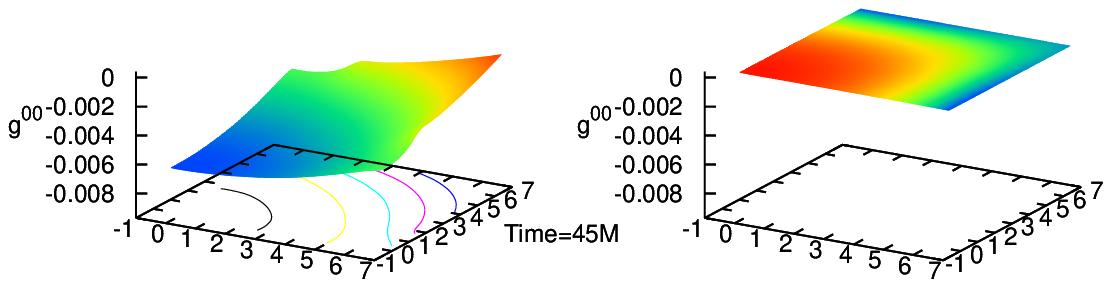}
      \caption{The $tt$ component of the metric for a Brill wave of
        amplitude $a=0.5$, comparing constraint-preserving boundary
        conditions with the standard Sommerfeld conditions.  The above
        plot shows a two-dimensional cut in the xy plane at various
        times.  On the right is the evolution of the brill wave with
        constraint-preserving SAT and on the left is the same
        simulation but with standard Sommerfeld type boundary
        conditions.\label{fig:brill_slides}}
  \end{center}
\end{figure}

As a first test of the implementation of the constraint preserving
boundary conditions for the full Einstein equations, we have
considered low amplitude wave solutions of the linearized Einstein
system.  These solutions exhibit non-trivial dynamics which exercise
the boundaries, but for which the source terms of the Einstein
equations are negligible. The particular initial data which we use are
the quadrupole Teukolsky waves~\cite{Teukolsky82}, which have been
used as a testbed in a number of numerical
studies\cite{Rinne2007,Rezzolla97a,Abrahams97a,Rezzolla99a}.  The
particular solution which we use follows Eppley~\cite{Eppley79} in
combining in-going and outgoing wave packets so as to produce a
solution which is regular everywhere in the space-time.

The overall behaviour of the evolutions using our three boundary
conditions is summarized in Fig.~\ref{fig:teukolsky}, which plots the
evolution of the $L_2$-norm of the Hamiltonian constraint as a
function of coordinate time, for a wave of amplitude $0.001$.  In each
case, there is a reduction of the constraint violation as the wave
propagates off the grid. In the standard Sommerfeld case, this quickly
saturates at a level of $10^{-7}$, determined by the finite
differencing resolution. In the case of the SAT boundary conditions,
however, the constraint violation eventually reaches machine round-off
due to the constraint damping in the interior of the domain. This
happens at a much faster rate for the explicitly constraint preserving
condition (``CP-SAT'') which introduces the modification described in
Sec.~\ref{sec:cpbc}. It is notable that in this case, the initial
boundary reflection which the standard Sommerfeld condition shares
with the simple SAT treatment, is also absent.

\subsection{Nonlinear waves}
\label{sec:brill}

The goal of our boundary treatment is to reduce the errors introduced
into the evolution domain during evolutions of strong field space-times
involving non-linear waves, as for instance, generated during
binary black hole evolutions. To model this problem in a simplified
setting which does not involve complications due to excision or
interior mesh-refinement boundaries, we have carried out tests
using the nonlinear Brill wave solutions~\cite{Brill59}. These
solutions have been studied in a number of numerical contexts,
both as testbeds, as well as exploring the onset of black hole
formation~\cite{Eppley77, Eppley79,Holz93, Alcubierre99b, Pazos2006}.
The initial spatial metric takes the form
\begin{equation}
  ds^2 = \Psi^4 [ e^{2q} (d\rho^2 + dz^2) + \rho^2 d\phi^2],
\end{equation}
in cylindrical $(\rho, \phi, z)$ coordinates. We choose $q$ of the
form of a Gaussian packet centered at the origin,
\begin{equation}
  q = a \rho^2 e^{-r^2},
\end{equation}
where $a$ is a parameter which is used to set the overall amplitude
of the axisymmetric wave. Generally we choose a value of $a=0.5$
to construct a wave which is strong, but not so as to evolve to
a black hole. As a result, we expect the initially nonlinear
solution generate waves which propagate off the grid leaving
behind a flat space-time.

\begin{figure}
  \begin{center}
    \resizebox{85mm}{!}{
      \includegraphics{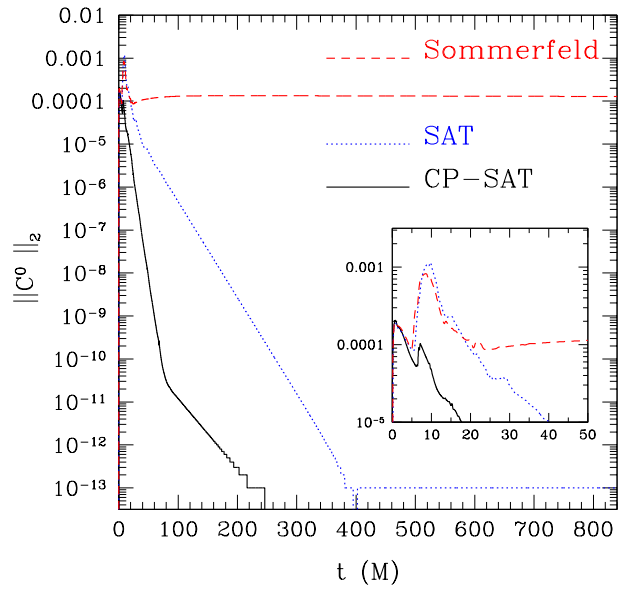}
    }
    \caption{The $L_2$-norm of the harmonic constraints for a Brill wave
      of amplitude $0.5$, comparing constraint-preserving boundary
      conditions with the standard Sommerfeld conditions, as well as the
      purely Sommerfeld SAT algorithm to ensure
      well-posedness. \label{fig:brill_c0}}
  \end{center}
\end{figure}

In Fig.~\ref{fig:brill_slides} we show a number of frames from two
evolutions, displaying the metric $\gamma^{tt}$ component at various
time instances on a grid $7$ units in size. In the right column, the
standard Sommerfeld conditions have been used, whereas on the left we
have used the constraint preserving SAT boundary conditions. By the
second frame at $t=8$, the wave pulse has reached the boundary, and
the following frames show the reflected pulse. Qualitatively, the
CP-SAT boundary conditions show a much smoother profile, with smaller
amplitude features. By $t=45$, the wave has left the grid in the CPSBP
case, to the extent that it cannot be seen on the linear scale of the
figure. In the standard Sommerfeld case, however, there is still some
non-trivial dynamical evolution.  A more quantitative demonstration is
shown in Fig.~\ref{fig:brill_c0}, which plots the $L_2$-norm of the
harmonic constraint $C^0$ as a function of coordinate time for three
situations: The standard Sommerfeld boundary conditions
(``Sommerfeld''), the SAT boundary conditions developed in
Sec.~\ref{sec:bc} (``SAT''), and the constrained version of these
boundary conditions, following the prescription of Sec.~\ref{sec:cpbc}
(``CP-SAT''). In the Sommerfeld case, the constraint violation is
entirely reflected by the grid boundaries, and the value remains
essentially constant at its initial value throughout the evolution of
the evolution, even though constraint damping has been used on the
interior code. The SAT boundary conditions, however, do a much better
job of removing constraint violation from the grid, showing the
exponential decrease with time that is expected from the damped
solution. The constraint preserving boundary conditions show the
strongest damping, suggesting that the constraint violating modes
introduced by these boundary conditions are much smaller than for the
SAT case. The evolution of the other constraint components show the
same behaviour.
\begin{figure}
  \begin{center}
    \resizebox{85mm}{!}{
      \includegraphics{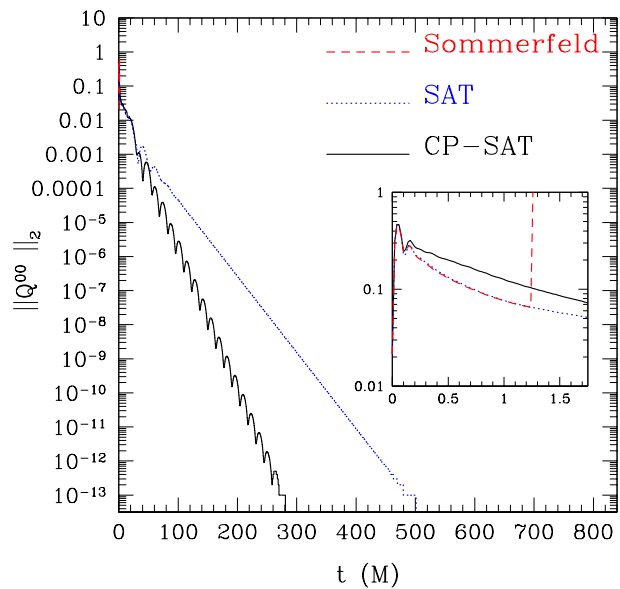}
    }
    \caption{Evolution of $Q^{00}$ component of the harmonic data for a
      Brill wave perturbed by random noise of a kernel amplitude of
      $\epsilon \pm 0.075$, over all the grid points.  This is placed on
      top of Brill wave initial data with an amplitude of
      $a=0.5$. \label{fig:brill_rand}}
  \end{center}
\end{figure}

\begin{figure}
  \begin{center}
    \resizebox{85mm}{!}{
      \includegraphics{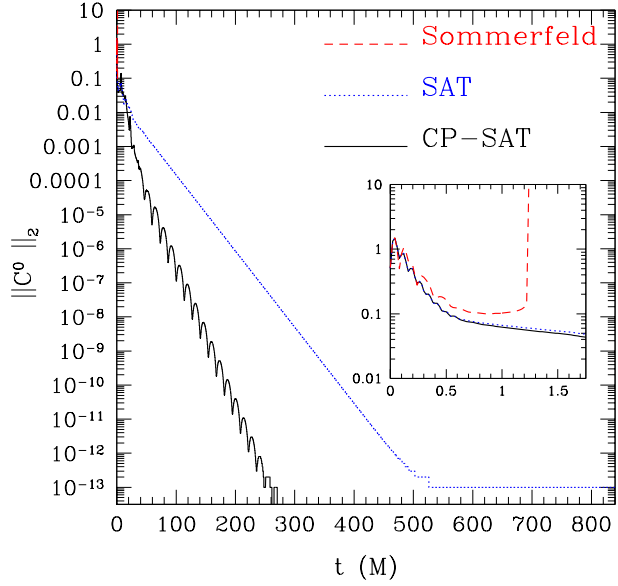}
    }
  \end{center}
    \caption{Evolution of the $L_2$-norm of the harmonic constraints for
      a Brill wave ($a=0.5$) perturbed by a checkerboard noise pattern of
      amplitude $\epsilon \pm 0.1$, over all the grid points, in order to
      excite the highest frequency grid mode. \label{fig:brill_check}}
\end{figure}

As a final test of the stability of our boundary prescription, we have
carried out evolutions of Brill waves for which we have attempted to
excite high-frequency error modes along the lines of the ``robust
stability'' test~\cite{Alcubierre:2003pc, Szilagyi00a}.  This test is
a means of determining whether it is possible for modes of any
frequency within any of the grid variables to exhibit exponential
growth during the evolution. On a numerical grid, error modes exist at
fixed frequencies, set by the grid resolution, and the standard test
consists of perturbing each variable at each grid point by a small
amount of randomly determined amplitude $\epsilon$. The effect of the
random perturbation is to seed modes which then have the potential to
grow, if the system is unstable at that frequency. Since being first
used in~\cite{Szilagyi00a} and proposed as a standard testbed
in~\cite{Alcubierre:2003pc}, the test has been used in a number of
applications to demonstrate well-posedness of numerical
implementations~\cite{applesweb1,Alcubierre2003:mexico-I,
  rinne-2006-23, Szilagyi00a, Szilagyi02b, Calabrese:2005ft}.  In
Fig.~\ref{fig:brill_rand} we applied this test by applying some kernel
of random data to all points including the boundary points.  For the
SAT methods the random noise gets damped and then the decay of the
energy looks similar to that of the standard brill test in
Fig.~\ref{fig:brill_c0}. For the standard Sommerfeld boundary
conditions the evolution becomes unstable at the boundaries.

A variant of this test recognizes that in the case of an ill-posed
system, the fastest exponential growth will result from the highest
frequency mode. On a finite-difference grid, the frequency of this
mode is set by the grid spacing. We can excite this mode by adding
perturbations to the data in a ``checkerboard'' pattern, where
neighboring points receive an opposite perturbation of fixed amplitude
$\epsilon$. That is, we choose
\begin{equation}
  \epsilon_{ijk} = \left\{
    \begin{array}{ll}
      +\epsilon, & \textrm{for $i+j+k$ even}, \\
      -\epsilon, & \textrm{for $i+j+k$ odd.}
  \end{array}
  \right.
  \label{eq:checkerboard}
\end{equation}

In Fig.~\ref{fig:brill_check} we show the evolution of the $L_2$-norm of
the $C^0$ constraint component for the evolution of an $a=0.5$ Brill wave
for which each component of the initial data has been modified according
to Eq.~(\ref{eq:checkerboard}) with $\epsilon=0.1$.  The two versions of
the SAT boundary conditions prove to be rather impervious to the initial
data perturbation, and display essentially the same behaviour as in the
unperturbed case, Fig.~\ref{fig:brill_c0}.  It is perhaps notable that
the non-constraint-persevering boundary conditions show a slightly
slower decay rate than for the non-perturbed data of
Fig.~\ref{fig:brill_c0}, so that it takes more than $100$ time units to
reach the level of machine round-off, whereas the constraint preserving
boundary conditions reach this level in essentially the same amount of
time as in the unperturbed case (though with a somewhat different decay
profile). The simple Sommerfeld boundary conditions, however, are unable
to cope with the initial perturbation and lead to an instability on a
very short timescale.

\section{Conclusions}
\label{sec:conclusions}
We have examined the initial boundary value problem for the
second-order formulation of the Einstein equations in the generalized
harmonic gauge.  The system of evolution equations for this
finite-difference harmonic code was derived in~\cite{Szilagyi:2006qy}
where it was shown to be accurate, stable, and convergent for
long-term evolutions of black hole space-times, such as head-on
collisions of two black holes, isolated black holes, and binary black
hole inspiral and merger.  In this paper we described the derivation,
implementation and testing of a new boundary treatment for this
system.  We demonstrated that this new treatment maintained the
validity and convergence (to lower order) seen with the standard
boundary treatments. We additionally show that these conditions give
us greater accuracy (for all reasonable resolutions), improved
constraint preservation, improved boundary transparency, and greater
stability in robust stability tests.

We implemented Sommerfeld-type boundary conditions as in
Eq.(\ref{eq:sommerfeld}), which are applied via the simultaneous
approximation term (SAT) method to control the energy growth of the
system, and are designed to be maximally dissipative.  We then
establish well-posedness for the semi-discrete symmetric hyperbolic
evolution system via the energy method~\cite{Carpenter1999a} by
bounding the energy growth of the system under the assumption that the
boundaries are in the linearized regime.  We have implemented
finite-differencing stencils that obey the summation by parts (SBP)
rule~\cite{Strand1994a} with the diagonal norm, with minimum bandwidth
second-derivative SBP stencils as derived in~\cite{Carpenter1994a}.
These stencils give fourth-order accuracy in the interior, and
second-order at the boundary.  While the standard stencils give
fourth-order everywhere, we show that the improved accuracy of the SBP
conditions more than makes up for the loss of two orders of
convergence.

The stability and well-posedness of the boundary conditions has been
demonstrated for a number of test problems: shifted scalar waves,
linearized waves, nonlinear waves, and random and high frequency
stability tests. Further improved accuracy results from incorporating
the constraint preservation into the conditions, following the
prescription of~\cite{Kreiss:2006mi, Babiuc:2007rr}.  The boundary
conditions are still Sommerfeld type for most metric components, but
we substitute conditions gained from enforced preservation of the
harmonic constraints.  This gives us four conditions directly from the
harmonic constraints, three from the coupling of these conditions to
our outgoing Sommerfeld-type conditions, and the three components for
the directions tangent to each boundary face come only from our
Sommerfeld-type conditions. In Sec.~\ref{sec:results} we show that, as
expected, these new outgoing Sommerfeld, constraint-preserving
conditions retain the robust stability and convergence properties of
the purely Sommerfeld-SBP conditions.  The tests also demonstrate that
these new conditions lead to smaller errors in satisfying the
constraints, and are more transparent to waves propagating through the
boundaries. They should thus lead to more accurate evolutions than the
purely Sommerfeld SBP penalty boundary conditions.

In a related study, Rinne et al.~\cite{Rinne2007} have considered a number
of boundary treatments for the case of a first-order in space harmonic
formulation, including the Kreiss-Winicour~\cite{Kreiss:2006mi} treatment
adopted here for a second-order system. They find that an additional
physically motivated condition, $\partial_t \Psi_0=0$, which aims to
eliminate incoming radiation, can have important effects in reducing
physical reflections. Similar modifications may also prove beneficial
to the second-order system presented here, though apparent reflections
from the outer boundary are rather small even in the case of
non-linear waves studied in Sec.~\ref{sec:brill}.

With binary black hole evolutions now extending over multiple orbits,
and thus many crossing times on conventional computational grids,
boundary effects can potentially have a non-trivial influence on the
late-time dynamics and extracted gravitational wave signals from such
simulations.  The tests provided here, including nonlinear Brill wave
evolutions, suggest that these methods will also be effective for
isolated strong sources, and thus will also be appropriate for black
hole space-times, though these involve a number of other technical
considerations (such as excision) which we do not explore here. The
methods can be extended to other formulations of the Einstein
equations, provided certain hyperbolicity assumptions are satisfied,
and we are currently pursuing improvements of other commonly used
systems such as the conformal-traceless one employed
in~\cite{Pollney:2007ss}.

\ack 
It is a pleasure to to thank Michaela Chirvasa and Tilman Vogel for
enlightening conversations regarding well-posedness and summation by
parts. Partial support to this research comes through the SFB-TR7
``Gravitationswellenastronomie'' of the German DFG. The numerical
results presented here in this paper were obtained using the clusters at
the AEI.



\section*{References}

\bibliographystyle{iopart-num}

\bibliography{aeireferences}

\appendix
\section*{Appendix}
\label{sec:appendix}
\setcounter{section}{1}

As an instructive example which contains the essential features of the
derivation for the Einstein equations, we derive explicitly the energy
estimate for the wave equation with shift, 
\begin{equation}
  \partial_{t}^{2}u = \left( \frac{-\gamma^{ij}}{\gamma^{tt}}\partial_{i}\partial_{j} 
  - 2\frac{\gamma ^{it}}{\gamma^{tt}}\partial_{i}\partial_{t}\right)u.
\label{eq:shwave}
\end{equation}
where $-\frac{\gamma ^{it}}{\gamma^{tt}}$ is the shift $\beta^{i},$
and $\beta^{i}\beta^{j} - \frac{\gamma^{ij}}{\gamma^{tt}}$ is the lapse.

We need to ensure that the energy, $\mathcal{E}^{\left(n\right)} =
\Vert u\left(\cdot,t\right)\Vert^{2},$ satisfies that the energy of
the system is bounded for the duration of the simulation. The time
derivative of the energy of the system can be re-written in
semi-discrete form as follows:
\begin{eqnarray}
  \frac{d}{dt}\mathcal{E} & = & \frac{d}{dt}\left( \Vert{u_{t}} \Vert^2 + 
    \left\Vert{-\frac{\gamma^{ij}}{\gamma^{tt}}u_{i}u_{j}}\right\Vert \right)  \nonumber\\
  & = &
  (\langle u_{t}, u_{tt} \rangle + \langle u_{tt}, u_{t} \rangle)
  -\frac{\gamma^{ij}}{\gamma^{tt}}(\langle u_{i}, u_{jt} \rangle 
  + \langle u_{it}, u_{j} \rangle)\, .
\label{eq:energy}
\end{eqnarray}
In this section our notation will follow that: we will use partial
derivative symbols for continuum equations and subscripts for
semi-discrete derivatives.  To ensure that this quantity remains bounded
in the semi-discrete case, we determine the energy growth which arises
from the application of our boundary conditions, and remove this via
the simultaneous approximation term (SAT, or ``penalty'')
method~\cite{Carpenter1994a}. We use a discrete second derivative
stencil which also obeys SBP and more accurately approximates a second
derivative than the wide stencil created from applying our first
derivative twice.

Since we use differencing operators which obey the SBP condition, we can
make use of Eq. (\ref{eq:sbp}) to integrate Eq. (\ref{eq:energy}).  For the
wave equation, after some algebra, this condition gives
\begin{equation}
  \frac{d}{dt}\mathcal{E} = -2\left[\left.\frac{\gamma^{ij}}{\gamma^{tt}}(u_{t}u_{j})\right\vert_{x_{i}=0}^{x_{i}=N_{i}}
  + \left.\frac{\gamma^{it}}{\gamma^{tt}}(u_{t}^{2})\right\vert_{x_{i}=0}^{x_{i}=N_{i}}\right] \, .
\label{eq:estimate}
\end{equation}
That is, the change in energy is determined by fluxes at the boundary
points, $x_{i}=0$ and $x_{i}=N_{i}$. 

On the boundary faces, we impose a set of conditions which for
the moment we write in a generic form
\begin{eqnarray}
  \left[\beta_{x_{i}=0}\partial_{t} + \alpha_{x_{i}=0}\partial_{i}
  + \delta_{x_{i}=0}\right] \left( u - u_{0} \right) &=& 0
  \label{eq:bc1} \\
  \left[ \beta_{x_{i} = N}\partial_{t} - \alpha_{x_{i}=N}\partial_{i}
  - \delta_{x_{i}=N}\right] \left( u - u_{0} \right) &=& 0
  \label{eq:bc2}
\end{eqnarray}
in terms of parameters $\alpha$, $\beta$, and $\delta$ which are
indexed according to the grid face. These are substituted into
into the estimate, Eq. (\ref{eq:estimate}), leading to
\begin{equation}
\fl
\qquad \qquad
  \frac{d}{dt}\mathcal{E}= 
  -2\left[\left.\left(\frac{\alpha_{N_{i}}}{\beta_{N_{i}}}u_{t}^{2} -
    \frac{\gamma^{it}}{\gamma^{tt}}u_{t}^{2}\right)\right\vert_{x_{i}=N_{i}} 
  - \left.\left(\frac{\alpha_{0_{i}}}{\beta_{0_{i}}}u_{t}^{2} - \frac{\gamma^{it}}{\gamma^{tt}}u_{t}^{2}\right)\right\vert_{x_{i}=0}\right]\,,
\label{eq:penalty1}
\end{equation}
where $\eta^{i}$ is the normal to the boundary face $i$, and $u_0$
are data chosen to be consistent with the initial data.
The SAT method allows us to choose values for the free parameters
in the boundary terms which conserve the energy in the
system. We first write the original shifted wave equation,
Eq.~(\ref{eq:shwave}), in semi-discrete form, explicitly including
the boundary terms:
\begin{eqnarray}
\fl
&& \hskip 1.0cm u_{tt}  =  -\frac{\gamma^{ij}}{\gamma^{tt}}H^{-1}D^{(2)}_{ij}u
  - 2\frac{\gamma^{it}}{\gamma^{tt}}H^{-1}D^{(1)}_{i}u_{t}
+ \tau_{0_{i}}H^{-1}E_{0_{i}}(\alpha_{0_{i}}u_{t}+\beta_{0_{i}}S_{i}u + \delta_{0_{i}}u)
  \nonumber\\
\fl
 & & \hskip 1.90cm + \tau_{N_{i}}H^{-1}E_{N_{i}}(\alpha_{N_{i}}u_{t}+\beta_{N_{i}}S_{i}u + \delta_{N_{i}}u)\, .
\label{eq:penalty2}
\end{eqnarray}
The $E_{a}$ are vectors of length $N$ defined as $E_{N_{i}}=
(0,0\ldots 0,1)^{\top}$ and $E_{0_{i}}= (1,0\ldots ,0)^{\top}$ to be
zero everywhere except at the boundary points. $S_{i}$ are
sideways blended finite differencing stencils satisfying the SBP
property, as described in the previous section.

We determine the time dependence of the energy for this new system in
order to derive coefficients $\tau$ for our penalty terms which give a
well-posed semi-discrete system. Substituting Eq.~(\ref{eq:penalty2})
into Eq.~(\ref{eq:energy}), and once again making use of the SBP
property, Eq.~(\ref{eq:sbp}), we arrive at
\begin{eqnarray}
\fl
& & \hskip 1.0cm \frac{d}{dt}\mathcal{E} = \left(\tau_{N_{i}}\alpha_{N_{i}} -
  \frac{\gamma^{it}}{\gamma^{tt}}\right)u_{t}^{\top}E_{N_{i}}u_{t} +
  2\left(\tau_{0_{i}}\alpha_{0_{i}} + \frac{\gamma^{it}}{\gamma^{tt}}\right)
  u_{t}^{\top}E_{0_{i}}u_{t} 
  \nonumber\\
\fl
  & & \hskip 1.90cm +2\left(\tau_{N_{i}}\beta_{N_{i}} -
  \frac{\gamma^{ij}}{\gamma^{tt}}\right)u_{t}^{\top}E_{N_{i}}S_{i}u 
  +2\left(\tau_{0_{i}}\beta_{0_{i}} +
  \frac{\gamma^{ij}}{\gamma^{tt}}\right)u_{t}^{\top}E_{0_{i}}S_{i}u\, .
\end{eqnarray}
The free parameters $\tau_0$ and $\tau_N$ can be used to eliminate
the $u_{t}^{\top}E_{N_{i}}S_{i}u$ terms, by setting
$\tau_{0}\beta_{0}= -\gamma^{ij}/\gamma^{tt}$ and
$\tau_{N}\beta_{N}=\gamma^{ij}/\gamma^{tt}$. Then, the energy
evolves according to
\begin{eqnarray}
  \frac{d}{dt}\mathcal{E} & = &
  -2\left(\beta_{N_{i}}\frac{\gamma^{it}}{\gamma^{tt}}-\alpha_{N_{i}}
    \frac{\gamma^{ij}}{\gamma^{tt}}\right)\beta_{N_{i}}^{-1}u_{t}^{\top}
    E_{N_{i}}u_{t}  \nonumber\\
    && + 2\left(\beta_{0_{i}}\frac{\gamma^{it}}{\gamma^{tt}}-\alpha_{0_{i}}
    \frac{\gamma^{ij}}{\gamma^{tt}}\right)\beta_{0_{i}}^{-1}u_{t}^{\top}
    E_{0_{i}}u_{t} = 0\,.
\end{eqnarray}
The last equality is arrived at after some algebra, substituting the
boundary conditions, Eq.~(\ref{eq:bc1}--\ref{eq:bc2}), and making
use of the original wave equation, Eq.~(\ref{eq:shwave}).

The resulting semi-discrete evolution equation is given by
\begin{eqnarray}
\fl
\qquad
   u_{tt} & = & -\frac{\gamma^{it}}{\gamma^{tt}}H^{-1}D^{(1)}_{i}u_{t}
  - \frac{\gamma^{ij}}{\gamma^{tt}}H^{-1}D^{(2)}_{ij}u 
  - \frac{\gamma^{ij}}{\gamma^{tt}\beta_{0_{i}}}H^{-1}E_{0_{i}}
  (\alpha_{0_{i}}u_{t} + \beta_{0_{i}}S_{i}u  + \delta_{0_{i}}u) 
  \nonumber\\
\fl
\qquad
  && +\frac{\gamma^{ij}}
  {\gamma^{tt}\beta_{N_{i}}}H^{-1}E_{N_{i}}(\alpha_{N_{i}}u_{t}
  +\beta_{N_{i}}S_{i}u + \delta_{0_{i}}u )\, ,
  \label{eq:sat}
\end{eqnarray}
which, as a result of the application of the SAT terms, satisfies
the energy conservation equation $d\mathcal{E}/dt = 0$.

\end{document}